\documentclass[12pt]{article}

\usepackage{amssymb}
\usepackage{amsfonts}
\usepackage{amsmath}

\def\Z{\mathbb Z}

\def\R{\mathbb R}
\def\NN{\mathbb N}

\newcommand{\+}{{\dagger}}
\newcommand{\sfrac}[2]{{\textstyle\frac{#1}{#2}}}

\newcommand{\ep}{{\mathrm{e}}}
\newcommand{\diff}{{\mathrm{d}}}

\renewcommand{\and}{{\quad{\rm and}\quad}}

\renewcommand{\=}{\ =\ }

\advance \textheight by \topskip
\makeatletter
\@addtoreset{equation}{section}

\makeatother

\begin{document}

\begin{titlepage}
\setcounter{page}{0}
\begin{flushright}
ITP--UH--17/15\\
\end{flushright}

\vskip 2.0 cm

\begin{center}

{\LARGE\bf
A Calogero formulation for \\[8pt]
four-dimensional black-hole micro states
}

\vspace{20mm}

{\Large Olaf Lechtenfeld${}^{\times}$ \ and \
Suresh Nampuri${}^{\+}$ 
}
\\[10mm]

\noindent ${}^\times${\em
Institut f\"ur Theoretische Physik and Riemann Center for Geometry and Physics\\
Leibniz Universit\"at Hannover \\
Appelstra\ss{}e 2, 30167 Hannover, Germany }\\
{Email: olaf.lechtenfeld@itp.uni-hannover.de}\\[10mm]

\noindent ${}^\+${\em
CAMGSD-IST\\
Universidade de Lisboa\\
Av. Rovisco Pais, 1049-001, Lisbon, Portugal}\\
{Email: nampuri@gmail.com}\\[10mm]

\vspace{20mm}

\begin{abstract}
\noindent
We extract the leading-order entropy of a four-dimensional extremal black hole in ${\cal  N}{=}2$ ungauged supergravity by  formulating  the CFT$_1$ that is holographically dual to its near-horizon  AdS$_2$ geometry, in terms of a rational Calogero model with a known counting formula for the degeneracy of states in its Hilbert space.
\end{abstract}

\end{center}

\end{titlepage}

\newpage

\section{Introduction}
A successful statistical mechanical description of black-hole microstates constitutes  one of the most precise tests of any purported theory of quantum gravity such as string theory. The most outstanding insight to be gleaned from string theory can be formulated in terms of the holographic AdS/CFT correspondence which establishes an isomorphism between the Hilbert space of quantum gravity in asymptotically AdS spaces and that of a conformal field theory living on the lower-dimensional boundary of the AdS space. Hence, non-perturbative objects in gravity such as black holes have a microstate description as thermal ensembles in the holographically dual theory. The least well understood of the well-studied AdS/CFT correspondences is the AdS$_2$/CFT$_1$ pair, where the dual conformal quantum mechanics is still an outstanding formulation problem in string theory. AdS$_2$ is of more interest than just as a two-dimensional toy model of quantum gravity: Every extremal black hole in four dimensions possesses a near-horizon geometry that can be expressed as the direct product of a black hole in  AdS$_2$ and a spherical, planar or hyperbolic horizon of the four-dimensional black hole. The deep-throat geometry of the AdS isolates the constant modes in it from the asymptotic modes of fields in the black-hole background that affect the black-hole horizon and hence its entropy. In fact, the constant modes in the near-horizon geometry are fixed in terms of the quantum numbers of the black hole, and they are independent of their asymptotic  values. This is the well known attractor mechanism displayed by these extremal black holes (see \cite{David:2006yn} and \cite{Sen:2008yk} and references therein for a detailed explication). The holographic Bekenstein--Hawking entropy of the black hole is therefore determined purely by states in the near-horizon region. Hence, an encoding of these states in the dual conformal quantum mechanics attains significance in identifying the holographically dual conformal quantum mechanics and counting the microstates of the black hole.
In this article, we look at the induced worldline superconformal quantum mechanics of an $n$-particle BPS system moving in the background of a black-hole in AdS$_2$. This quantum mechanics has a reformulation~\cite{Gibbons:1998fa} in terms of an $n$-particle rational Calogero model (of type $A_{n-1}$),\footnote{
See also \cite{Galajinsky:2012vh, Galajinsky:2015apa} for recent related work.} 
and we argue that this encodes the thermal-ensemble states corresponding to the black hole in the holographically dual CFT$_1$.  We justify this assertion by counting the large-charge degeneracy of states in this model to arrive at the Bekenstein--Hawking entropy of the dual black hole in AdS$_2$.

\section{Calogero dynamics and extremal black holes}
The near-horizon geometry of a zero-temperature BPS black-hole solution in four-dimensional ungauged supergravity  is a black hole in AdS$_2 \times S^2$, whose geometry is described as 
\begin{equation} \label{BH}
\diff s^2 \= - \sfrac{r^2 - \Delta^2}{b_*^2} \diff t^2 + \sfrac{b_*^2}{r^2- \Delta^2} \diff r^2 
+ b_*^2 \diff\Omega_2^2  \ .
\end{equation}
We restrict ourselves to only bosonic backgrounds in the theory. The scalar fields ${\phi^i}$ that make up the moduli space in this background and do not correspond to flat directions of the scalar potential are driven to a critical point of this potential. They flow from the asymptotically flat space to the near- horizon geometry, and their extremum values ${\phi_*^i}$ are fixed entirely in terms of the quantum numbers of the system, independent of the asymptotic starting values. Hence, the near-horizon geometry acts as an attractor in the moduli space.
The common radius of the AdS$_2$ and $S^2$ spaces is the modulus $|Z|$ of the central charge~$Z$ of the supersymmetry algebra and, by the BPS condition, equal to the mass $M(\phi^i)$ of the black hole. Both are computed at a point in the asymptotic moduli space coinciding with the attractor point. The three U-duality invariants characterizing the black hole can hence be summarized as
\begin{equation}
M(\phi_*) \= |Z(\phi_*)| \= b_* \ .
\end{equation}

As our model system, we consider a bound state of D0 and D4 branes wrapped on a $ CY_3 \times T^2$ to produce a four-dimensional dyonic black solution. In the M-theory picture, this can be viewed as a collection of particle momenta on the M-theory circle $S^1_M$ with intersecting M5 branes wrapping a (4-cycle in $CY_3  ) \times S^1_M$. As the near-horizon geometry decouples from the 
asymptotically flat space, the states contributing to the black-hole entropy must be localized in this region. Hence, probing the Hilbert space of these states will yield a count of the black-hole microstates from a statistical mechanics perspective. As mentioned in the previous section, the Hilbert space of quantum gravity in the  near-horizon AdS$_2$  geometry can be formulated in terms of states in the holographically dual CFT$_1$ which implies that  the black-hole degeneracy must be reproducible in terms of the counting formula for states in this conformal quantum mechanics. We therefore need a proposal for identifying the microstates of the AdS$_2$ black hole in a conformal quantum mechanical theory. One such proposal for conducting such an analysis is motivated by the observation that this system belongs to the special class of BPS black holes which can be lifted up to five dimensions to yield a near-horizon geometry of a BTZ black hole in AdS$_3 \times S^2$. 
The holographic correspondence with the two-dimensional BCFT is well understood in this case, and the black hole can be thought of as a chiral-ensemble excitation in the CFT, with the central charge defined by the D4 branes and with the CFT excitation number of the black hole being equal to its mass. Hence, in the 'black hole in AdS$_2$' scenario, we are motivated to consider the black hole as an excitation about AdS$_2$, described in terms of degrees of freedom that can be encoded in terms of a superconformal quantum mechanics. This suggests that the black hole is naturally represented as a halo of $n$~BPS~particles moving in the AdS$_2$ background. These particles are governed by a superconformal quantum mechanics, with a target space that is the symmetric product of AdS$_2$ and $S^2$. This is a putative formulation of the holographically dual CFT. We proceed to delineate this connection below.

\section{AdS$_2$-Calogero correspondence} 
\subsection{Rational Calogero from AdS$_2$}
 Gravity in two dimensions is a conformal quantum field theory living on a strip. States in this theory are in a one-to-one correspondence to those defined in the BCFT, which in this case is also the holographically dual field theory. This field theory is in fact some superconformal quantum mechanics and must encode all the bulk states. A single particle moving in the AdS background is described by a  superconformal quantum mechanical worldline theory. For a scalar particle, in the large-radius limit of an AdS geometry, parametrized in the Poincar\'e patch via
\begin{equation}
\diff s^2 \= - \frac{R^4}{q^4} \diff t^2 + R^2 \frac{\diff q^2}{q^2} \  , 
\end{equation}
this is the rational 2-body Calogero model, with the Hamiltonian\footnote{
See \cite{Gibbons:1998fa} and \cite{Claus:1998ts} for a detailed exposition.}
\begin{equation}
H \= \frac{p^2}{2 } + \frac{\lambda^2}{q^2}\ ,
\end{equation}
where $\lambda$ is proportional to the angular momentum Casimir of the particle in four dimensions.
The energy must be evaluated with respect to the AdS global time coordinate, where the Killing vector is smooth everywhere, and the Hamiltonian for this coordinate is given by 
\begin{equation}\label{H}
H \= \frac{p^2}{2 } + \frac{\lambda^2}{q^2}+ \omega^2\frac{q^2}{2} \ ,
\end{equation}
with an undetermined non-zero force constant~$\omega$.
The addition of the last term arises by passing from  the Poincar\'e time~$t$ to the global time~$\tau$, which are related as 
\begin{equation}
\partial_\tau \=\partial_t + \omega^2 K\ ,
\end{equation}
where $K$ is the special conformal transformation generator of the SO(2,1) isometry group of AdS$_2$, given by $K = \frac{1}{2}q^2$ in the large-$R$ limit. The ground-state wave function in this case reads
\begin{equation}
\psi(q)\= q^\alpha\, \ep^{- \omega^2 q^2/4}
\qquad\textrm{where}\qquad
\alpha = \sfrac12 \bigl(1 + \sqrt{1 + 4 \lambda^2 }\bigr) \ .
\end{equation}
Hence, the particle has no support at the center of AdS, and its wave function is localized farther out.  The limiting value of the wave function at the boundary acts as a local insertion on the BCFT and, hence, defines the operator in the BCFT corresponding to some state in the bulk. As a consequence, a state corresponding to an excitation in AdS$_2$ can be mapped to a superparticle moving in the bulk and such states can be organized in terms of the asymptotic symmetry group of AdS$_2$. 
Thus, we can regard the black hole as an ensemble of $n$~BPS particles in  AdS, which define a superconformal quantum mechanics with a target space given by $n$ symmetrized copies of AdS$_2 \times S^2$. In the fully symmetric sector, the SU(2) R-charge of the superconformal quantum mechanics will be simply the common R-charge of the $n$~particles multiplied by $n$. It follows that the angular momentum matrix of this system is a multiple of the identity matrix.

Quantizing the spectrum of this system will generate the Hilbert space that counts the entropy of  the BPS black hole. 
To this end, we observe that, in our chosen model of the dyonic black hole as a supersymmetric D0-D4 bound-state ensemble, the microstate counting is essentially a field-theory computation of the Witten index for $n$~particles. Their momenta are equal to the $D0$-brane quantum numbers in the two-dimensional worldvolume theory of intersecting M5 branes on $CY_3 \times S^1_M$, at a point in the moduli space corresponding to $V_{CY_3} \ll R_{S^1_M}$. This theory is simply two-dimensional SU($n$) super Yang-Mills on a cylinder,\footnote{
See \cite{Gibbons:1998fa} and references therein for details.}
which has been shown in~\cite{Gorsky:1993pe} to be equivalent to an $n$-particle rational Calogero model governed by the Hamiltonian
\begin{equation}
H \= \sum_i \frac{p_i^2}{2 }\ +\ \sum_{i<j}\frac{\lambda^2}{ (q_i-q_j)^2}  \ .
\end{equation}
As in the single-particle case, the spectrum of the system is computed with respect to the global time~$\tau$, and the corresponding Hamiltonian can be related to the Schwarzschild-time Hamiltonian by adding the superconformal generator~$K$. 
This introduces a confining harmonic  well to the rational Calogero model, 
\begin{equation}\label{CC}
H \= \sum_i \frac{p_i^2}{2}\ +\ \sum_{i<j} \frac{\lambda^2}{(q_i - q_j)^2}
\ +\ \omega^2 \sum_i \frac{q_i^2}{2}\ .
\end{equation}
In the Higgs limit where the spacing between the positions of all particles vanish and all the particles are driven to the origin of the coordinate system, the analysis of the ground states is similar to that of the single-particle system, and so the discussion for the single-particle case goes through for the multi-particle system.
Hence, this model offers a putative formulation of the CFT$_1$ required to count the large-charge leading-order black-hole entropy. We now proceed to show how this model encodes the vacuum states of the holographically dual quantum mechanics and how the AdS$_2$ geometry emerges in the bulk by analyzing the flow of the ground state in the space of its coupling constants. We test this model by deriving the degeneracy formula for this system.

\subsection{AdS$_2$ from Calogero}
How may an asymptotically AdS$_2$ bulk background arise from a rational Calogero model? It is necessary to check that the ground state of this model in some limit (corresponding to approaching the boundary, i.e.~$q\rightarrow 0$) must move in the space of coupling constants of the deformed model, such that this state feels the vacuum geometry of the bulk gravity theory, namely AdS$_2$.
The metric it should see is nothing but the Fischer information metric for the ground-state wave function of the deformed Calogero Hamiltonian~(\ref{H}), given by 
\begin{equation}
\psi(q, \alpha, \omega) \= A \,q^\alpha e^{-\omega^2 q^2/2}
\qquad\textrm{where}\quad 
A^2 \= (2\alpha{+}1) (\sfrac{\omega}{2})^{2\alpha + 1}\ .
\end{equation}
In the limit of $q\rightarrow 0$, the wave function can be approximated to leading order by 
\begin{equation}
\psi(q, \alpha, \omega)\=A \,q^\alpha
\qquad\textrm{for}\quad 
0 \leq q \leq \sfrac{2}{\omega}\ ,
\end{equation} 
while all higher-order deformations of the Hamiltonian can be neglected. 
This yields a two-parameter space graded by $\alpha$ and $\omega$. The Fischer information metric for a space parametrized by $n$~variables $\Theta=(\theta_i)$, with $i=1,\ldots,n$, is given as 
\begin{equation}
g_{\theta_i \theta_j} \=\int_{0}^{\frac{2}{\omega}} \!\diff{q}\ q^{2\alpha}\ 
\partial_{\theta_i} \log |\psi(q,\Theta)|^2\
\partial_{\theta_j} \log |\psi(q,\Theta)|^2\ |\psi(q,\Theta)|^2\ ,
\end{equation}
where $|\psi(q,\Theta)|^2$ is the probability density on the wave-function space. 
The Fischer metric on the two-dimensional space under consideration is computed explicitly to be 
\begin{equation}
\diff s^2\= 
\tilde{\alpha}^{-2}\bigl( \diff\tilde{\alpha}^2 + \omega^{-2}\diff\omega^2 \bigr)
\qquad\textrm{where}\quad \tilde{\alpha}=\sfrac1{2\alpha{+}1}\ .
\end{equation}
Hence, to summarize, if one considers a deformation of the conformal Calogero Hamiltonian by a harmonic oscillator term which initiates a flow in the space of coupling constants, then in the limit of $q\rightarrow 0$, to observe the change in the ground state, we need to consider only the quadratic deformation so as to obtain a two-dimensional space of coupling constants. The latter is found to be essentially Euclidean AdS$_2$.\footnote{
As $\alpha > \frac{1}{2}$, this is not a complete Poincar\'e patch, since $0\le\tilde{\alpha}<\frac12$. However, in what follows, we will simply refer to it as the Poincar\'e patch and leave the inherent subtleties in this metric for future study.} 
This explicitly goes to show that the flow of the ground state in the space of relevant coupling constants, near the boundary of AdS$_2$, falls into a representation of the SL($2,\R$) symmetry group that annihilates the vacuum of the dual CFT$_1$.  
Hence, we now have a dynamical model which is a putative candidate for counting  the degrees of freedom of the  holographically dual CFT$_1$. We now run our first check of  this counting by computing the degeneracy of  states in  the spectrum of this Hamiltonian, dual to the ground state of a BPS particle moving in the background of a black hole in AdS$_2$. 

\section{Degeneracy from the Calogero Hamiltonian}
The presence of the harmonic oscillator discretizes the $n$-body spectrum in (\ref{CC}) so that it acquires energy eigenvalues~\cite{poly,felepo},
\begin{equation}
E_n (m)\= \omega \bigl(f(\lambda) + \sfrac{n}{2} + m\bigr) \qquad\textrm{with}\quad m\in\Z_{\ge0}\ .
\end{equation}
In the above, $f(\lambda)$ is a linear function of $\lambda$. Here, the quantum number~$m$ is actually partitioned into positive-integer parts of maximum size~$n$,
\begin{equation}
m \= m_1 + m_2 + m_3 + \ldots \qquad\textrm{with}\quad m_r\in\{1,2,\ldots,n\}\ ,
\end{equation}
which determines the multiplicity of $E_n(m)$ to be the number $p_n(m)$ of correspondingly restricted partitions of~$m$.
Its generating function reads~\cite{cole}
\begin{equation}
\sum_m p_n(m)\,q^m \= \prod_{1\leq k\leq n} \frac{1}{(1- q^k)}
\qquad\textrm{with}\quad q= \ep^{-\beta}\ .
\end{equation} 
Here, $\beta$ is the periodicty of the Euclidean time circle and (upto, numerical factors) equal to the inverse of the black-hole temperature.
We work in the large-$n$ limit, which implies $p_n(m)\to p(m)$ and simplifies the generating function to 
\begin{equation}
\sum_m p(m)\,\ep^{-\beta m}\=\prod_{k\in\NN} \frac{1}{(1- q^k)}
\=\frac{1}{\eta(\beta)} \ep^{\frac{\pi \beta}{24}}\ .
\end{equation}
The asymptotic growth of $p(m)$ can be obtained by a saddle-point approximation of the Laplace transform of the degeneracy formula, in the low $\beta$ limit, and by using the transformation property of the Dedekind $\eta$ function under Poisson resummation to give 
\begin{equation}
p(m) \ \approx\ \ep^{2 \pi \sqrt{\frac{m}{6}}}\ ,
\end{equation} 
where the approximation sign indicates a suppression of all quadratic corrections to the saddle point and of other subleading terms.
As the system we are studying exhibits no classical mass gap,\footnote{
This is completely consistent with looking at the most symmetric sector of the theory.} 
we need to pick the largest possible Euclidean time periodicity to define the Euclidean temperature, and hence we take the Euclidean periodicity to be $\frac{2 \pi n}{\omega}$. This is equivalent to rescaling $\omega$ in the spectrum by a factor of $n$ and demanding that we count only eigenvalues with $m$ being integral multiples of~$n$.
Therefore, in the above expression, $m$ should actually be replaced by $mn$. 

Now, let us consider the physically relevant values of this model for 
the black-hole statistical mechanics. Essentially, we are counting a Witten index on the full Hilbert space of the system, and so we should be looking at the ground-state degeneracy. The full conformal quantum gravity has a net central charge of zero, which is the sum of the conformal anomalies due to diffeomorphisms, ghosts and matter. As the matter content in the black-hole background does not differ from that of 'empty' AdS, the matter contribution to the stress tensor is the same in both cases, and hence the only matter contribution to the stress tensor can come from modes which are fully annihilated by the complete SO(2,1) isometry of the AdS$_2$ vacuum. This fixes the excitation quantum number to
\begin{equation}
m \= \frac{c}{24} \ .
\end{equation}  

Another argument for the above relation can be put forward as follows. The ground-state degeneracy we are counting is in the black-hole background, while the Calogero spectrum has been evaluated in the Poincar\'e patch of AdS$_2$. A conformal transformation can be used to map the ground state of the black-hole background to that of the Poincar\'e patch. Under this transformation the stress tensor picks up an inhomogenous term coming from the Schwartzian derivative, which raises the 
 ground-state energy by an amount of $\frac{c}{24}$ in the Poincar\'e patch~\cite{Cadoni:2000gm}).  Here, $c$ is the ground-state Casimir energy or central charge of the holographically dual CFT. 
From a dual CFT perspective, this implies that all such black holes must have a Casimir energy equal to $\frac{c}{24}$, implying again that $m = \frac{c}{24}$.

The number~$n$ of particles in the Calogero model is equal to the number of degrees of 
freedom of the CFT$_1$ and thus equal to~$c$. Consequently, $mn = \frac{c^2}{24}$, and the leading-order contribution to the black-hole entropy is found to be 
\begin{equation}
S \=  2\pi \sqrt{\sfrac{c^2}{6\times 24}} \= 2 \pi \frac{c}{12} \ , 
\end{equation}
which matches the standard  Bekenstein--Hawking black-hole entropy of the four-dimensional black hole reduced on the two-sphere in the near-horizon geometry~\cite{Castro:2008ms}. Note that the relevant degrees of freedom that go into the computation of the black-hole entropy can be interpreted as the degrees of freedom of the AdS vacuum that the black-hole observer does not see, resulting in an entanglement entropy. 
Hence, one can extract leading-order information about the microstate description of bulk states in AdS by using general properties of an equivalent formulation of the BCFT in terms of a known superconformal Calogero model.

\section{Discussion and conclusions}
The formulation of a holographic dual to quantum gravity in AdS$_2$ has been the least well understood of the frequently analyzed gauge-gravity correspondences.  Concurrently, extremal black holes in four dimensions with a near-horizon geometry have a density of states that is related to the square root of the energy, reflecting an underlying degeneracy of microstates that is captured by a CFT$_2$ as opposed to a CFT$_1$. This article builds upon a proposed formulation in \cite{Gibbons:1998fa} of the microstates of a black hole in AdS$_2$ in terms of the worldline quantum mechanics of conformal Calogero particles in AdS$_2$. The degeneracy of states in this model accurately reproduces the Bekenstein--Hawking entropy without taking recourse to viewing the underlying CFT as the chiral half of a two-dimensional CFT or implementing the Cardy formula. The accuracy of the computation indicates that this formulation offers a putative way to understand quantum gravity in AdS$_2$ and opens avenues for new checks on the gauge-gravity correspondence in two dimension. 

If the Calogero model is to be dual to string theory in AdS$_2$, then the metric on the space of coupling constants, as generated by the flow of generic states in this space, must be the emergent bulk metric of the geometry in which the motion of a BPS particle is governed by the worldline Hamiltonian that includes those couplings. The background so derived is dual to the states whose flow is under consideration. We have already demonstrated this for the vacuum state as a necessary condition for this theory to be a holographic candidate for AdS$_2$. 
Investigating the Fisher information metric on the full Hilbert space of the Calogero model, by which bulk geometry emerges from superconformal quantum mechanics, might yield further insights into gauge-gravity duality in two as well as in higher dimensions.

\section*{Acknowledgements}
S.N.\ gladly acknowledges stimulating conversations and useful collaborations (which inspired some of the ideas here) with Vishnu Jejjala, Kevin Goldstein, Gabriel Cardoso, Michael Haack, Marco Zagermann, Cornelius Schmidt-Colinet and Paolo Benincasa. S.N.\ would also like to thank the 
Riemann Fellowship awarded by the Riemann Center for Geometry and Physics which initiated this collaboration, and the FCT fellowship FCT - DFRH - Bolsa SFRH/BPD/101955/2014 
under which his research was conducted.
This article is also based upon work from COST Action MP1405 QSPACE, 
supported by COST (European Cooperation in Science and Technology).

\end{document}